\documentclass[preprint,aps,showpacs]{revtex4}

\usepackage{amsmath}
\usepackage{graphics}
\usepackage{graphicx}
\usepackage{epsfig}

\begin{document}

\title{Macroscopic quantum information channel via the polarization-sensitive interaction between the light and spin
subsystems}
\author{O.S. Mishina, D.V. Kupriyanov,}
\affiliation{Department of Theoretical Physics, State Polytechnic
University, 195251, St.-Petersburg, Russia}%
\email{ksana@quark.stu.neva.ru}%
\author{E.S.Polzik}
\affiliation{QUANTOP - Danish Quantum Optics Center, Niels Bohr
Institute, 2100 Copenhagen, Denmark}%
\email{polzik@phys.au.dk}%

\begin{abstract}
We discuss the quantum information channel, which is based on
coherent and polarization sensitive interaction of light and
atomic spin waves. We show that the joint Heisenberg dynamics of
the polarization Stokes components of light and of the angular
momenta of atoms has a wave nature and can be properly described
in terms of the macroscopic polariton-type spin wave created in
the sample. The principles of the quantum memory and readout
protocols via the wave coupling of the output time modes in the
light subsystem and the output spatial modes in the spin subsystem
are demonstrated.
\end{abstract}

\maketitle%

\section{Introduction}
Quantum information processing based on continuous variables as an
alternative to the discrete quantum schemes has been intensely
studied during the last decade, see, for example, the review
\cite{SaBPtVL} and reference therein. Spin oriented atomic
ensemble and polarized coherent light can be considered as
convenient physical objects for storage and for transport of the
quantum information, which can be written in the fluctuations of
their polarizations. There are two experimental demonstrations of
the entanglement and memory protocols in Refs.\cite{JKP,JSFCP},
where the dynamical coupling of the spin fluctuations of an atomic
ensemble consisting of cesium atoms with the polarization
fluctuations of linearly polarized light via Faraday effect were
implemented. The main feature and certain advantage of those
protocols are in that only collective integral variables of light
and of atomic subsystems contributed in the quantum fluctuation
interchange. However, as was recently reported in
Ref.\cite{KMSJP}, in a more general situation the light and spin
subsystems develop the polariton-type spin wave dynamics in the
sample. That means that the dynamics of the Stokes polarization
components and of the atomic spins becomes coupled not integrally
and collectively but locally in time as well as in space during
the whole interaction cycle. In this case the quantum correlations
are spread among all the possible time and spatial polariton modes
created in the sample. In this report we further develop the ideas
of Ref.\cite{KMSJP} and show how they could be applied for
organizing the readout and memory protocols based on the wave
dynamics of the process.

\section{Macroscopic polariton-type spin wave}
Consider an off-resonant pulse of radiation probing the
spin-polarized atomic ensemble during the short time interval so
that incoherent scattering is negligible. Then, as shown in
\cite{KMSJP}, the relevant Heisenberg operators describing the
light and atomic subsystems, and responsible for the quantum
correlations dynamics, obey the following wave-type equations
\begin{eqnarray}
\left[\frac{\partial}{\partial z}+\frac{1}{c}%
\frac{\partial}{\partial t}\right]%
\hat{\Xi}_1(z,t)&=&%
-\kappa_2\,\hat{\Xi}_2(z,t)\;+\;%
2\beta\,\bar{\Xi}_3\,%
\hat{{\cal J}}_z(z,t)%
\nonumber\\%
\left[\frac{\partial}{\partial z}+\frac{1}{c}%
\frac{\partial}{\partial t}\right]%
\hat{\Xi}_2(z,t)&=&%
\phantom{+}\kappa_2\,\hat{\Xi}_1(z,t)\;-\;%
2\epsilon\,\bar{\Xi}_3\,%
\hat{{\cal J}}_y(z,t)%
\nonumber\\%
\frac{\partial}{\partial t}\hat{{\cal J}}_z(z,t)&=&%
\phantom{+}\Omega\,\hat{{\cal J}}_y(z,t)\;-\;%
\epsilon\bar{{\cal J}}_x\,%
\hat{\Xi}_1(z,t)%
\nonumber\\%
\frac{\partial}{\partial t}\hat{{\cal J}}_y(z,t)&=&%
-\Omega\,\hat{{\cal J}}_z(z,t)\;+\;%
\beta\bar{{\cal J}}_x\,%
\hat{\Xi}_2(z,t)%
\label{1.1}%
\end{eqnarray}%
In these equations the light pulse with 100\% linear polarization
along $x$-direction propagates through the sample along
$z$-direction. The atoms also have 100\% orientation of their
angular momenta along the $x$-direction. The field subsystem is
described by the Heisenberg operators for two polarization Stokes
components $\hat{\Xi}_1(z,t)$ and $\hat{\Xi}_2(z,t)$. The first
Stokes component $\hat{\Xi}_1(z,t)$ is responsible for imbalance
between photon fluxes of the modes linearly polarized along $\xi$
and $\eta$ axes rotated with respect to $x$ and $y$ directions by
$\pi/4$ angle, and the second component $\hat{\Xi}_2(z,t)$ is
responsible for imbalance in the right-hand and the left-hand
polarizations. The third Stokes component $\bar{\Xi}_3$
responsible for imbalance in $x$- and $y$-type polarizations is a
quantum integral of motion and considered as external
(non-operator) parameter in equations (\ref{1.1}). The atomic
subsystem is described by the operators $\hat{\cal J}_{z}(z,t)$
and $\hat{\cal J}_{y}(z,t)$, which are the Heisenberg operators
for the spatial distribution of the transverse fluctuations of the
angular momenta in respect to $z$ and $y$ axes. The spin
projection $\bar{\cal J}_{x}$ on $x$ direction is another quantum
integral of motion and is also considered a c-number parameter.

The system of equations (\ref{1.1}) describes the coupled
wave-type dynamics of the Stokes components of the light subsystem
and of the transverse fluctuations of the atomic angular momenta.
Since the field subsystem consists of many photons and the spin
subsystem consists of macroscopic number of individual atomic
spins, we shall call such a wave as a macroscopic polariton-type
spin wave. This polariton wave describes the mesoscopically smooth
dynamics for the interacting microscopic quantum variables of the
light and spin subsystems. Such a wave represents a manifestly
quantum object, which exists only at the level of quantum
description of the field and spin fluctuations. While deriving
equations (\ref{1.1}) it was assumed that atoms preserve their
location during the interaction cycle and there is no destruction
of the coherent wave dynamics coming from the random atomic
motion. This important condition can be fulfilled, for example,
for interaction duration longer than tenth of milliseconds for an
ensemble consisting of ultracold atoms. Alternatively for hot
atoms in a gas cell the interaction time should be shorter than a
few microseconds.

The following two important parameters govern the dynamics of the
process. The first one is $\beta$, which is the angle of the
polarization rotation of the probe light due to Faraday effect per
one spin flip in the ensemble in $z$-direction. The second is
$\epsilon$, which is the ellipticity induced in the propagating
light by the atomic sample due to Cotton-Mouton effect per one
spin flip in $y$-direction . Frequency $\Omega=\Omega_0+\Omega_2$
combines the regular precession caused by the external magnetic
field $\Omega_0$ with the frequency of light-induced shift of the
Zeeman sublevels $\Omega_2$. Parameter $\kappa_2$ is responsible
for birefringence effects with respect to $x$- and $y$-type
polarizations of the probe, i.e. for the unitary transformation of
linear polarization to circular polarization and vice versa.

In general the system of equations (\ref{1.1}) can be solved only
numerically. But in a special case it can be solved analytically
if at least one of the parameters, either $\Omega$ or $\kappa_2$
approaches zero. In practice this can be happened if, for example,
the magnetic field compensates the contribution of the light shift
so that $\Omega_0=-\Omega_2$. The solution of the system
(\ref{1.1}) for this special case will be discuss in details
elsewhere. Here we restrict our further discussion to an even
simpler model where $\kappa_2=0$ and $\Omega=0$ simultaneously.
Although it is not trivial to compensate the difference in the
refractive indices for $x$- or $y$-type polarized probe
propagating in the spin polarized atomic ensemble, it is very
appealing to consider this particular case, due to the following
two reasons. 1) There are no quantum correlations under the above
conditions induced between the field and atomic subsystems due to
regular spin precession and due to average birefringence of the
sample. 2) Such an approximation lets us clear identify the basic
features of the macroscopic spin polariton dynamics and their
potential importance for the respective quantum information
channel. In addition, as a technical simplification, we will
assume that the propagation time through the sample is negligibly
small, and the retardation effects are unimportant in the
frequency domain we are going to discuss.

\section{Entanglement between the Laplace modes}

In experiment the light pulse probes an atomic sample of the
length $L$ during the finite time $T$. However, because the wave
dynamics is developing only in the forward direction in space as
well as in time, we can formally extend this process up to
infinite interaction time and consider the probe light propagating
in a semi-infinite medium. Then parameter $L$ can be associated
with a selected layer in such a medium located at the coordinate
$z=L$ and after interaction time $T$ the state of the system is
considered at the moment $t=T$. Such an extension lets us define
the following Laplace modes for the field and for the atomic
subsystem
\begin{eqnarray}
\hat{\Xi}_i^{\rm{out}}(s)&=&\hat{\Xi}_i(L,s)=\int_{0}^{\infty}dt\,%
e^{-st}\,\hat{\Xi}_i(L,t)%
\nonumber\\%
\hat{{\cal J}}_{\mu}^{\rm{out}}(p)&=&\hat{{\cal J}}_{\mu}(p,T)=%
\int_{0}^{\infty}dz\,%
e^{-pz}\,\hat{{\cal J}}_{\mu}(z,T)%
\label{2.1}%
\end{eqnarray}%
for $i=1,2$ and $\mu=z,y$ respectively. In the above
transformations we consider the outgoing field operators at point
$z=L$ at any time and define the Laplace $s$-mode for them. The
outgoing atomic operators are considered at a selected moment of
time $t=T$ but for an arbitrary spatial location and are described
in terms of the spatial Laplace $p$-mode.

The solution of the system (\ref{1.1}) can be conveniently
rewritten using relations between the Laplace modes in the
following form
\begin{eqnarray}
e^{-p(s)L}\,\hat{\Xi}^{\rm{out}}_1(s)\;%
-\;\frac{2\beta\bar{\Xi}_3}s\,%
e^{-sT}\,\hat{{\cal J}}^{\rm{out}}_z(p(s))&=&%
\nonumber\\%
&&\hspace{-3.5cm}\int_{T}^{\infty}dt\,e^{-st}\,\hat{\Xi}^{\rm{in}}_1(t)\;%
-\;\frac{2\beta\bar{\Xi}_3}s\,%
\int_{L}^{\infty}dz\,e^{-p(s)z}\,\hat{{\cal J}}^{\rm{in}}_z(z)%
\nonumber\\%
e^{-p(s)L}\,\hat{\Xi}^{\rm{out}}_2(s)\,%
+\frac{2\epsilon\bar{\Xi}_3}s\,%
e^{-sT}\,\hat{{\cal J}}^{\rm{out}}_y(p(s))&=&%
\nonumber\\%
&&\hspace{-3.5cm}\int_{T}^{\infty}dt\,e^{-st}\,\hat{\Xi}^{\rm{in}}_2(t)\;%
+\;\frac{2\epsilon\bar{\Xi}_3}s\,%
\int_{L}^{\infty}dz\,e^{-p(s)z}\,\hat{{\cal J}}^{\rm{in}}_y(z)%
\label{2.2}%
\end{eqnarray}
The initial Heisenerg operators $\hat{\Xi}^{\rm{in}}_i(t)$ and
$\hat{{\cal J}}^{\rm{in}}_{\mu}(z)$ of the field and atomic
subsystems contribute on the right hand side and are defined in
their original form as functions on time and spatial coordinates.
There are two additional relations, not shown here, which together
with the relations (\ref{2.2}) lead to the full solution in the
Laplace form. The main feature of the solution (\ref{2.2}) is that
$s$- and $p$- Laplace modes are not independent one on another but
coupled by the dispersion relation caused by the wave nature of
the process
\begin{eqnarray}
p&=&p(s)\;=\;\frac{A}s,
\nonumber\\%
A\;&=&\;-2\beta\,\epsilon\,\bar{\Xi}_3\,\bar{\cal J}_x
\label{2.3}%
\end{eqnarray}
And because of complete symmetry of the problem similar dispersion
relation can be written for $s=s(p)$.

As clearly seen from the solution (\ref{2.2}) in an extended
medium $L\to\infty$ and after long interaction time $T\to\infty$
there will be entanglement developed between the temporal and
spatial dynamics inside the polariton wave mode. When the field
and atomic subsystems are separated the mode entanglement would
manifest itself in either entanglement or swapping of the outgoing
quantum states. The wave nature of such type of quantum
correlations will be visualized after a certain spectral selection
in the light subsystem and a spatial spectral selection in the
spin subsystem are made.

The important parameter which determines the type of the output
quantum correlations is the polariton group velocity. To define
this velocity the inverse Laplace transformation should be written
in terms of a Fourier integral, with the following parametrization
of the Laplace modes: $s=-i\omega$ and $p=iq$. Then the transport
dynamics of the correlation wave is characterized by the group
velocity
\begin{equation}%
v_{g}\;=\;\frac{d\omega}{dq}\;=\;-\frac{A}{q^{2}}%
\label{2.4}%
\end{equation}%
As one can see the group velocity can be either positive or
negative. The latter case appears when the product
$\beta\epsilon<0$, as, in example of alkali atoms, can occur for
the probe in the blue wing of $D_2$-line. For such a specific
situation there is no stationary point in the corresponding
wavepacket expansion. The quantum fluctuation are exponentially
enhanced in space as well as in time with preserving their quantum
correlations. This case requires a special discussion. Below we
consider the alternative and a more typical for the standard wave
dynamics situation when the group velocity is positive and
$\beta\epsilon>0$. In case of alkali atoms this relates to the red
wing of $D_2$-line.

\section{Quantum memory and readout protocols}

Consider an experimental situation when a low-frequency mode with
the frequency $\omega$, such that $\omega T< 1$, is detected in
the output polarization state of the transmitted light. For this
mode, according to the dispersion law (\ref{2.3}) and (\ref{2.4}),
the group velocity can be quite low. As consequence the input
quantum state of the entire system will be transport through the
sample as a slow propagating wavepacket. Then the output
Heisenberg operators of the light subsystem will be actually
formed in the readout process of the input quantum operators of
atomic spins and preferably of those spatial components which are
distributed near respective high-frequency spatial mode
$q=A/\omega$.

\begin{figure}[tp]
\includegraphics{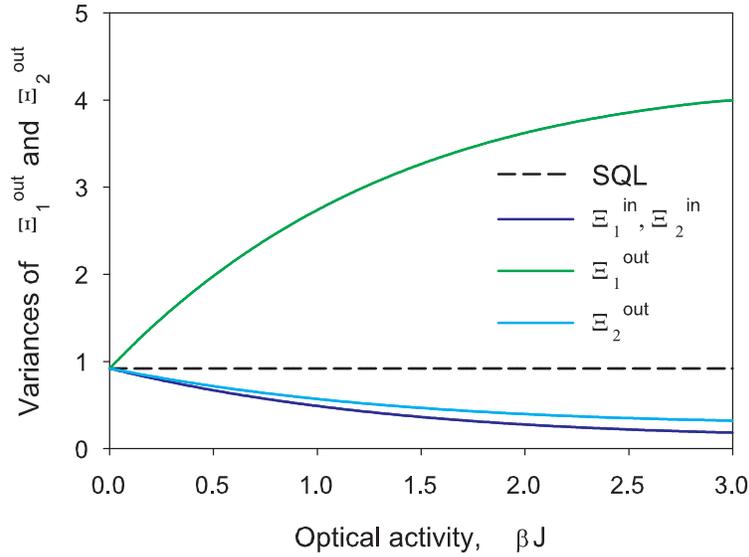}
\caption{The variances of the integral output Stokes components,
defined by Eq.(\ref{3.1}) for $\omega T=0.5$, as a function of the
optical activity of the sample $\beta J$. The dotted curve
indicates the vacuum noise (standard) quantum limit (SQL). The
blue curve indicates the contribution of the input Stokes
components.}
\label{fig1}%
\end{figure}

This can be demonstrated by the graphs plotted in figure
\ref{fig1}. All the dependencies are the expectation variances for
the following target integral observables
\begin{equation}
\hat{\Xi}^{\rm{out}}_i=\int_0^Tdt\,\cos{(\omega t)}\;%
\hat{\Xi}^{\rm{out}}_i(t)%
\label{3.1}%
\end{equation}%
for $i=1,2$. The calculations are performed for the red wing of
$D_2$-line of ${}^{133}$Cs, for detuning of the probe about
$-1200$ MHz from the $F_0=4\to F=5$ hyperfine transition. The
composition parameter $-ALT$ is varied from $0$ (left border of
the graph) to the level of $2$ (right border of the graph) so that
the chosen in our calculations detecting frequency $\omega T=0.5$
would be asymptotically coupled with the spatial mode with the
wave number $qL=4$, in the way explained in the previous section.
The input quantum states of the light and spin subsystems obey
Poissonian statistics. The blue curve shows how the input quantum
fluctuations of the Stokes components are mapped onto the output
state. As one can see for the extended medium with high optical
activity $\beta J$ (where $J=\bar{\cal J}_xL$ is the total angular
momentum of the sample) the contribution of the input field
fluctuations becomes negligible. The main impact on the output
variances is made by the input fluctuations of the spin subsystem.
It also follows from the graphs that the variance of
$\hat{\Xi}^{\rm{out}}_1$ component is much bigger than
$\hat{\Xi}^{\rm{out}}_2$ and the latter is below the standard
quantum deviation. This is a direct consequence of inequality
$\beta\gg \epsilon$, which is typical in the case of $D_2$
transition of alkali atoms. So the mapping of the state of light
is in this case combined with the squeezing operation on the
state. The output polarization state of the probe light in the
low-frequency domain of its fluctuation spectrum becomes squeezed
and the high-frequency spatial modes of the spin states
$\hat{J}_{z}^{\rm{in}}$ and $\hat{J}_{y}^{\rm{in}}$ are
respectively mapped into the integral modes of the
$\hat{\Xi}^{\rm{out}}_1$ and $\hat{\Xi}^{\rm{out}}_2$ Stokes
components.

Let us reverse the experimental situation and let the quantum
state of the spin subsystem be controlled by the high-frequency
polarization fluctuations of light, such that $\omega T> 1$. Then
the quantum memory protocol can be achieved and the quantum state
of these modes can be mapped into the output low-frequency spatial
modes of the spin subsystem. Indeed, in this case the transfer of
input correlations is fast because the polariton group velocity
$v_g$ is high enough. The original input quantum state of the
system will be transported out of the sample with the polariton
wave. Then the low-frequency spatial modes (with $qL< 1$) of the
output spin state are mainly formed via accumulation of the input
quantum fluctuations of the transmitted light. Because of complete
symmetry of the equations (\ref{1.1}) with respect to the light
and spin subsystems, the memory protocol can be visualized by the
same graphs shown in figure \ref{fig1} with the following change
in notation. The target observables should be now associated with
the spin subsystem
\begin{equation}
\hat{J}^{\rm{out}}_{\mu}=\int_0^Ldz\,\cos{(qz)}\;%
\hat{{\cal J}}^{\rm{out}}_{\mu}(z)%
\label{3.2}%
\end{equation}%
for $\mu=z,y$. The blue curve indicates the reduction of the input
state mapping onto these observables. The abscissa of the plot
should be associated with $\beta\bar{\Xi}_3T$, which is the angle
of the collective spin rotation if the probe light were circular
polarized. After change in notation the graphs of figure
\ref{fig1} show how the initial polarization quantum state of
light, which existed in the high-frequency domain of its
fluctuation spectrum, can be mapped into the integral output
quantum state of atomic spins. Asymptotically (for the parameters
used in our calculations) this results in the mode coupling
between $\omega T=4$ time mode of the Stokes variables and the
$qL=0.5$ spatial mode of atomic spins.

\section{Conclusion}

In this report we have discussed how interchange by the
polarization quantum states of the light and atomic spins could be
implemented via the polarization sensitive interaction between
these subsystems. The main result is the identification of the
spectral domains, where the quantum states of the input temporal
or spatial fluctuations could be transferred from one subsystem to
another. If the polarization fluctuations of the light and spin
subsystems carried the quantum information it would make possible
to create the quantum information protocols for writing in or
reading out this information. This specific quantum channel does
not exactly copy an original quantum state from one physical
carrier to another, but creates a new quantum state in the target
system related in a known way to the quantum state originally
existing in the source system. This can be done because of
multimode nature of interaction in the entire atoms-field system
and the protocols are related only to the specially selected
spectral domains. In a general case the input-output
transformations in the usual space-time representation are
expressed via the fundamental solution of the wave-type Heisenberg
equations (\ref{1.1}) in integral form, see \cite{KMSJP}. For
interaction limited in space as well as in time the idealized mode
description can only approximately approach the real interaction
process. Because of the multi-mode character of the input-output
transformation the fidelity, usually applied to describe the
quality of a quantum information protocol for relatively simple
systems, has to be revised in order to become applicable. This
work is in progress and will be published elsewhere.

\section*{Acknowledgments}

The work was supported by the Russian Foundation for Basic
Research (RFBR-05-02-16172-a) and by the European grant within
network COVAQIAL. O.S.M. would like to acknowledge the financial
support from the charity Foundation "Dinastia".

\end{document}